\documentstyle[psfig]{aa}                     

\def\magcir{\raise -2.truept\hbox{\rlap{\hbox{$\sim$}}\raise5.truept
\hbox{$>$}\ }}

  \title{Is 3C111, an apparently normal radio galaxy, the counterpart of 3EG J0416+3650?}
  \subtitle{}

   \author{ V. Sguera\inst{1}, L. Bassani\inst{2}, A. Malizia\inst{2}, A. J. Dean\inst{1}, R. Landi\inst{2},
  J. B. Stephen\inst{2}}

   \offprints{sguera@astro.soton.ac.uk}

   \institute{ School of Physics and Astronomy, University of Southampton, Highfield, Southampton, SO17 1BJ, UK
\and IASF/CNR, via Piero Gobetti 101, I-40129 Bologna, Italy}
   \date{Received / accepted}

  \titlerunning{}

   \authorrunning{Sguera et al.}

\begin{document}

\abstract{The Third EGRET Catalog (3EG) lists 66 high-confidence identifications 
of sources with Active Galactic Nuclei (AGN). All are classified as belonging to the blazar class, with the only
exception of the nearby radio galaxy Centaurus A. We report and strengthen the association of another 
radio galaxy, 3C111, with the EGRET source 3EG J0416+3650.
At the time of the compilation of the 3EG catalogue, 3C111 has been considered as a low-confidence counterpart of 3EGJ0416+3650, being
located outside the 99$\%$ $\gamma$-ray probability contour. Since this first suggestion, no other 
counterparts have been reported nor the EGRET error box has been searched for likely candidates. 3C111 has never been considerated
or cited in literature as a radiogalaxy counterpart of an EGRET source. We report a detailed multiwavelength 
study of the EGRET error box as well as for the first time the overall spectral energy distribution of 3C111, which appears to be
intriguingly similar to those of blazars, suggesting that the radiogalaxy 3C111 is the likely counterpart of 3EG J0416+3650.

\keywords{X-rays: observations- X-rays: individual(3EG J0416+3650)- Radiogalaxies objects: general}
}
 \maketitle


\section{Introduction}

The identification of sources detected by EGRET is one of the great challenges of current 
gamma-ray astronomy. More than sixty percent of the 271 high
energy gamma-ray sources reported in the 3rd EGRET Catalogue (Hartman
et al. 1999) are unidentified, with no firmly established counterparts
at other wavebands.  The majority of the identified objects are of extragalactic 
origin and belong to the blazar type. The 3EG catalogue lists 66 high-confidence 
identifications of blazars (labelled as ''A'') and 27 lower-confidence identifications (labelled as ''a'').
These objects have high and variable 
gamma-ray luminosities (often dominating the bolometric power) and power law spectra  
with an average  index of $\Gamma$=
2.15$\pm$0.04 (Mukherjee et al. 1997). Most of the EGRET  blazars are strong  radio sources (only a small fraction have a 
5GHz flux below 1 Jy) with flat spectra 
($\alpha$ $\le$ 0.5,  S$_{\nu}\propto\nu^{-\alpha}$). They also show significant 
optical polarization and/or variability in various wavebands; superluminal motion has 
been observed in some cases. \\ The combination of all these characteristics support the 
argument of relativistic beaming in jets which are pointing close to the line of sight.
In the widely adopted scenario of blazars, a single population of high-energy electrons 
in a relativistic jet radiate from the radio/FIR to the UV- soft X-ray by the synchrotron 
process and at higher frequencies by inverse Compton scattering soft-target photons present 
either in the jet (synchrotron self-Compton [SSC] model), in the surrounding ambient 
(external Compton [EC] model), or in both (Ghisellini et al. 1998 and references therein).
 Therefore  in the  blazar SED, two peaks corresponding to the synchrotron and inverse 
Compton components should be evident.\\
Very recently, attention has been focussed on the possibility that radio galaxies
could be also counterparts of unidentified EGRET objects. 
For example, Centaurus A (a Faranoff-Riley type I radiogalaxy) is the 
closest AGN (z=0.0018), and the only radio galaxy positively detected 
by EGRET (Sreekumar et al. 1999). Its gamma-ray spectrum is well characterized
by a single power law of photon index 2.4$\pm$0.28, steeper 
than the usual blazar-like $\gamma$-ray source, while the luminosity is $\sim$10$^{41}$ erg s$^{-1}$, 
about 10$^5$ time less  than that typically observed by EGRET. 
The nucleus of Centaurus A has a 5GHz  flux  greater than 1 Jy and  
a flat radio spectrum (Burns et al. 1983) plus 
a SED  similar to that of blazars (Chiaberge et al. 2001). Its jet is offset by an angle of $\sim$70$^{\circ}$
from the line of sight (Bailey et al. 1986, Fujisawa et al. 2000). 
There are three additional 
EGRET sources, for which a possible radio galaxy counterpart has been suggested. 
One such source is 3EG J1621+8203 (Mukherjee et al. 2002) for which NGC6251
(a Faranoff Riley type I radio galaxy) is the likely counterpart. 
Combi et al. (2003) have also recently reported the discovery of a double-sided 
Faranoff-Riley type II radio source, J1737-15, in the error box of 3EG J1735-1500; 
in this case, however,  the association with the EGRET source is more controversial 
due to the presence in the error box of another likely candidate.  
Also in these two cases, the EGRET spectrum is steeper and the gamma-ray luminosity 
lower than typically observed in blazars. Both sources have a flux at or around  5GHz 
lower than 1 Jy (0.36 Jy for NGC6251 while J1735-1500 is not detected), a nuclear 
radio spectrum which  is flat in NGC6251 (Jones  et al. 1986) and steep in 
J1735-15 (Combi et al. 2003); only in the case of  NGC6251 there are enough data in the 
literature  to show that  the nuclear 
SED is similar to blazars (Chiaberge et al. 2003). The radio jet of NGC6251 makes an angle of
45$^{\circ}$ with the line of sight (Sudou $\&$ Taniguchi 2000)
and the $\gamma$-ray luminosity  in this case is 3$\times$10$^{43}$ erg s$^{-1}$.
Compared with Cen A, the greater distance to NGC6251 could, perhaps, be compensated by a smaller angle between the jet and the line of sight.
More recently the possible association of the Seyfert 1 galaxy GRS J1736-2908 
with 3EG J1736-2908 has also  been proposed (Di Cocco et al. 2004). This AGN is not a strong radio emitter
having  a flux density at 5GHz and 1.4 GHz 
of respectively $\simeq$23 mJy and $\simeq$59 mJy and its radio spectrum is described by a power law with
$\alpha$=0.75, which however does not  fulfill the requirement 
of flatness ($\alpha$ $\le$ 0.5,  S$_{\nu}\propto\nu^{-\alpha}$).\\
To summarize the sparse observational evidence available at the moment indicate that 
if  radio galaxies are counterparts of EGRET sources, the 
characteristics are still similar to blazars but the source is of lower intensity; this
suggests the  possibility that we are dealing with objects where the jet is misoriented with respect to the 
observer. Recently Ghisellini, Tavecchio and Chiaberge (2004) have put forward a structured jet model
in which a slow jet layer is cospatial to a fast jet spine. 
In this model there will be a strong radiative interplay and feedback between the layer and spine as
both parts see extraphotons coming from the other part: this enhances the inverse compton emission
of both components thus explaining why also some radio galaxies can be relatively strong in $\gamma$-rays.\\
Here we present arguments in favour of the association of yet another radio  galaxy, 3C111,
with an EGRET source, 3EG J0416+3650. This result makes the probability   for radio galaxies 
to be counterparts  of unidentified EGRET sources even stronger and the search more 
promising given their much higher spatial density with respect to blazars. This could have 
important consequences in terms of the AGN  contribution to the 
gamma-ray background and the relation between aligned and misaligend blazars in 
terms of emission  mechanisms, AGN unified theory and jet physics.

\section{The gamma-ray source}
\begin{figure} [h]
\psfig{figure=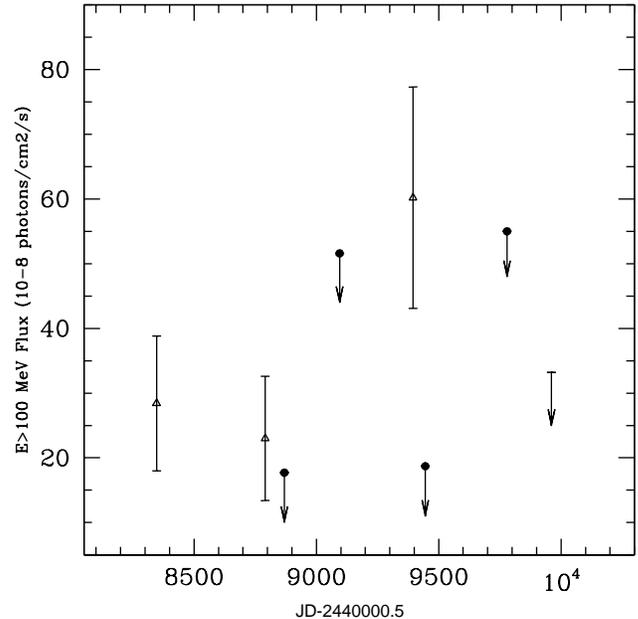,height=10cm,width=10cm}
\caption{Light curve of 3EG J0416+3650 from April 1991 to September 1995}
\end{figure}
3EG J0416+3650 is a gamma-ray source located at l=162$^{\circ}$.2 b=-9$^{\circ}$.97 (i.e. close to the galactic plane), 
with a 95${\%}$ confidence error radius of 38'.2 (Hartman et al. 1999). More recently Mattox al. (2001) 
have generated  elliptical fits to the 95$\%$ contours of the source: 
the semimajor and seminor  axes are  44'.6 and  32'.5 respectively with a 
position angle for the semimajor axis of 167$^\circ$. \\
In the third EGRET catalogue the source is labelled  as confused ("C") due to the presence 
of 4 nearby sources all having  gamma-ray fluxes greater than 3EG J0416+3650. The presence 
of these nearby sources can have an effect on the reported position and related uncertainty,
so that some allowance in the countour errors should be considered to account for 
this systematic 
effect (Hartman et al. 1999, section 5).\\
3EG J0416+3650 was not always bright enough to be detected 
by  EGRET. In fact the source does not appear in the second catalogue while in the third one
the source has been detected only in three Viewing Periods or VPs (VP=0.2+, from 1991 April 22 to 1991 May 07; VP=31.0, 
from 1992 June 11 to 1992 June 25; VP=321+, from 1994 Feb 08 to 1994 Feb 17); in the remaining VPs 
only upper limits to the flux were established.
In the cumulative exposure from multiple 
Viewing Periods (VP=1234, from 1991 April to 1995 October) the source is reported with a  significance of 5.3$\sigma$ and with 
an average  flux above 100 MeV of 1.3 10$^{-7}$ photon cm$^{-2}$ s$^{-1}$.
The gamma-ray spectrum is well fitted by a simple power law   with photon 
index $\Gamma$=2.59 $\pm$ 0.32
steeper than that generally observed in EGRET blazars and pulsars, which are by far 
the most obvious 
counterparts of unidentified gamma-ray sources.\\
The EGRET light curve from the
beginning to the end of the mission obtained from the 3EG catalogue (Hartman et al. 1999) 
suggests variability of the gamma-ray flux (see Fig.1). In fact Torres et al. (2001) have  
measured a variability parameter 
of I=2.61 for this source, slightly higher than the value of 2.5 which is 3 $\sigma$ away from 
the standard value of pulsars generally considered non variable. More recently Nolan et al. (2003) re-evaluated the 
variability of all EGRET sources by a different method and measured for 3EG J0416+3650 a 
variability index $\delta$ of 0.59. This value is significantly larger than the fractional
variability expected from steady sources, $\delta$ $<$ 0.14, thus confirming Torres et al. results.\\
Hartman et al. (1999) first suggested 3C111 (z=0.0485) as the optical counterpart of the EGRET 
source; but since the source optical position  is outside the 99$\%$ 
contour this radio galaxy  has been considered as a low confidence counterpart (type ''a'' in the 3EG catalog).  
Mattox et al (2001) estimated the a priory probability of association to be 0.2 and the 
a-posteriori to be 0.019, just slightly higher  than the minimum value (0.01) suggested 
for standard consideration of a potential identification. The low value obtained is  due 
to the large distance of 3C111 from the EGRET position ($\sim$76 arcmin) and to a radio spectrum  between
20 cm and 6 cm (Mattox et al. 2001) which did not fulfill the requirement of 
flatness. Since this first suggestion,  
no other counterparts have been reported 
nor the EGRET error box has been searched for likely candidates.
3C111 has never been considerated or cited in literature as a radio galaxy counterpart of an EGRET source.
\section{Search for X-ray/radio  counterparts in the EGRET error box}
As discussed above, the EGRET positional uncertainty  is large  and not 
always fully  reliable and this makes optical identification on the basis of position
 alone quite difficult. Other methods for identifying counterparts of EGRET objects have 
relied on information at other wavebands, particularly in the radio, looking for strong 
flat spectrum radio sources and/or  in X-rays,
searching for high energy emitting objects. Following this line, we have started a 
program to look for  high energy 
(above 10 keV) emitting sources located within EGRET error boxes via available high 
energy telescopes such as BeppoSAX 
(Sguera et al 2004) and INTEGRAL.
\begin{figure}
\psfig{figure=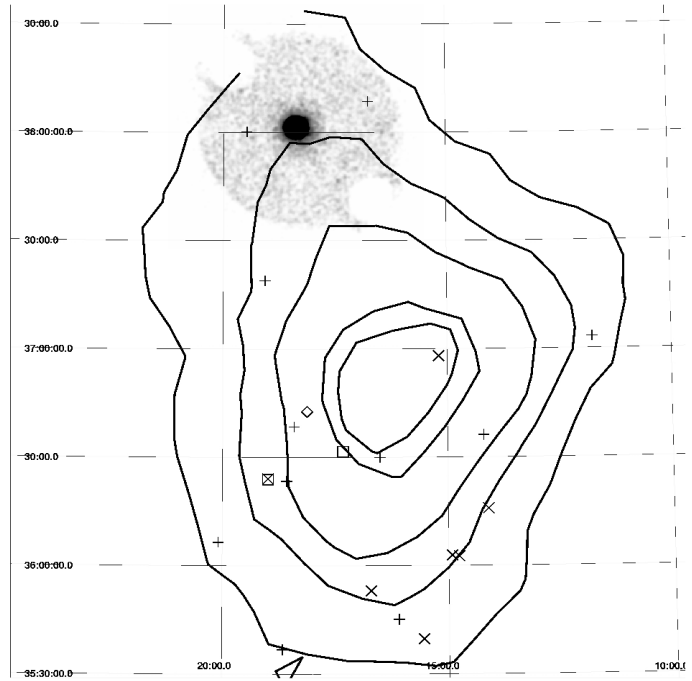,height=9cm,width=9cm}
\caption{The BeppoSAX-MECS (2-10 keV) image superimposed on the EGRET 
$\gamma$-ray probability contours at 50{\%}, 68{\%}, 95{\%}, 99{\%} and 99.9{\%}
confidence level.All  sources listed in table 1 and table 3 are  shown:
crosses are ROSAT Faint sources, diamond is Einstein Slew Survey source, pluses are NVSS radio sources and  squares are Green Bank radio sources.}
\end{figure}
In Fig. 2 the MECS (2-10 keV) image of the BeppoSAX observation targeted at 3C111
is superimposed on the EGRET $\gamma$-ray probability contours
(50{\%}, 68{\%}, 95{\%}, 99{\%} and 99.9{\%}). We note that 3C111 is located between the 99{\%} and 99.9{\%} contours.\\ 
The cross correlation of the 99.9{\%} EGRET error box with the Rosat All-Sky-Survey
catalogues (Bright B and Faint F) have resulted in 12 objects (1 and 11 respectively);
furthermore there is an Einstein Slew Survey source (ESS) 
which is not related to any of the Rosat sources . Therefore all together there 
are 13 X-ray sources within the EGRET  error box and they are shown in Fig.2 by means of different symbols. 
All these soft X-ray candidates
are listed in Table 1 together with their coordinates, their Rosat/Einstein
count rates and the offset from the 3EG source position. 
The only source reported in the Rosat Bright source catalogue is 3C111. All other
X-ray objects are either stars or unidentified in the Simbad/NED databases. Only one X-ray source, 3C111,  is
detected at  energies greater than a few keV.
Apart from this active galaxy, the unidentified Einstein Slew Survey source 1ES0414+365 (n.13 in Table 1)  
is the only other X-ray source in the field for which there might be reasons to consider it 
as a possible counterpart. It is a bright X-ray source (only a factor $\sim$1.4 dimmer than 3C111 in the 0.2-3.5 
keV), it is likely to be a transient (as was not observed by other X-ray instruments) and it is
the closest to the EGRET position (see Fig. 2). However, lacking information on its spectral shape (expecially above a few keV)
and nature, it is difficult at this stage to speculate further on its possible association to the EGRET source.\\
A cross correlation of all these X-ray objects with the radio catalogues available in the
HESARC database indicates that only one object is radio emitting: 3C111 (see Table 2 for details).  
To further investigate possible radio counterparts of the EGRET source, we have searched 
the NRAO/VLA sky survey (NVSS) catalog (Condon et al. 1998) for possible 1.4 GHz (20 cm) 
objects within the EGRET 99.9{\%} 
error contours. There are 8 sources with an integrated radio flux $\ge$ 100 mJy (see pluses in Fig. 2) 
and  they are all listed in Table 3
together with their coordinates, fluxes and the offsets from the 3EG source position.
The brightest one (NVSS J041820+380148,  n.8 in Table 3) 
\begin{table*} 
\begin{center}
\caption{X-ray sources in the EGRET error box.}
\begin{tabular}{llcccccc}
\hline
\hline
 & Source                   & RA            & Dec          & Count Rate   & Search Offset   & Type   & Radio Count. \\
 &                          & (J2000)       & J(2000)      & Cts/s        & $arcmin$          &        &              \\
\hline
1  & 1RXS J041821.6+380134  & 04 18 21.6     & +38 01 34.5    & 0.160     &  77.95            & B     & Yes\\  
2  & 1RXS J041630.5+362957  & 04 16 30.5    & +36 29 57.5   & 0.039    &  18.42            & F    &  No \\
3  & 1RXS J041412.2+363611  & 04 14 12.2     &+36 36 11.0    & 0.016    & 26.76            & F     &  No \\
4  & 1RXS J041824.5+363827  & 04 18 24.5     & +36 38 27.5    & 0.013    &  28.21            &   F    &  No \\
5  & 1RXS J041834.7+362325    & 04 18 34.7    & +36 23 25.5     & 0.022   &  37.74            &   F     &  No \\
6  & 1RXS J041903.8+371856     &04 19 03.8    & +37 18 56.5    & 0.010    &  46.18            &  F     &  No \\
7  & 1RXS J041145.6+370315   & 04 11 45.6     & +37 03 15.0      & 0.027   & 53.66           &  F     &   No  \\
8  & 1RXS J042005.2+360630   & 04 20 05.2     & +36 06 30.5    & 0.016   & 65.23            &  F     &    No \\
9 &  1RXS J041605.8+354510   & 04 16 05.8    & +35 45 10.5       & 0.019   & 65.60            &  F     &  No   \\
10 & 1RXS J041645.0+380837   & 04 16 45.0    & +38 08 37.5       & 0.027   & 78.20            &  F     &  No   \\
11 & 1RXS J041928.8+380010   & 04 19 28.8    & +38 00 10.5       & 0.014   & 80.14            &  F    &   No    \\
12 & 1RXS J041840.0+353649   & 04 18 40.0    & +35 36 49.0       & 0.016   & 80.15            &  F    &   No    \\  
13 & 1ES0414+365$^{\dagger}$  & 04 18 07.4   & +36 42 36.0  & 0.300     &  23.74    & ESS$^{\dagger}$    & No\\
\hline
\hline
\end{tabular}
\end{center}
Note: $\dagger$ = energy range counts rate is 0.2-3.5 keV \\
Note: $\dagger$ = Einstein Slew Survey
\end{table*}
\begin{table*}
\begin{center}
\caption{Radio counterparts of 3C111.}
\begin{tabular}{llcccccc}
\hline
\hline
 & Source                  & Flux   & Radio catalog \\
 &                         & mJy    &       \\
\hline
1  & NVSS J041820+380148   & 7726   & NRAO VLA Sky Survey (20 cm)   \\
2  & GB6J0418+3801         & 5168   & Green Bank 6 cm survey   (6 cm)  \\
3  & 0415+3754             & 6637   & 6cm Radio Catalog  (6 cm)    \\
4  & WN 0415.0+3754        & 12959  & Westerbork Sky Survey (92 cm)   \\
5  & TXS 0414+378          & 14017  & Texas Survey (82 cm)      \\
\hline
\hline
\end{tabular}
\end{center}
\end{table*}
\begin{table*}
\begin{center}
\caption{Radio sources in the EGRET error box.}
\begin{tabular}{llcccc}
\hline
\hline
 & Source                   & RA            & Dec          & Flux   & Search Offset   \\
 &                          & (J2000)       & J(2000)      & mJy        & $arcmin$         \\
\hline
1  & NVSS J041511+365802   & 04 15 11.2     & +36 58 02.8    & 260.8     & 15.78            \\  
2  & NVSS J041406+361548  & 04 14 06.4    & +36 15 48.4     & 110.8    &  40.89            \\ 
3  & NVSS J041859+362402  & 04 18 59.3    &+36 24 02.3     & 420.5   & 41.26           \\
4  & NVSS J041455+360250  & 04 14 55.1    & +36 02 50.7    &103.7    & 47.73            \\
5  & NVSS J041445+360236    & 04 14 45.9    & +36 02 36.2    &127.8   &  48.59            \\
6  & NVSS J041642+355303     &04 16 42.8    & +35 53 03.1    & 111.8    &  55.29          \\
7  & NVSS J041533+353940   & 04 15 33.2     & +35 39 40.7     & 218.9    & 71.42          \\
8  & NVSS J041820+380148   & 04 18 20.9     & +38 01 48.6     & 7726   & 75.85               \\
9  & GB6J0418+3801         & 04 18 22.1      & +38 01 47.0     &5168     & 75.93           \\
10 & GB6J0417+3631         & 04 17 19.6      & +36 31 36.0     & 123.0       & 24.10          \\
11 & GB6J0418+3623       & 04 18 59.4   & +36 23 53.0     & 112.0       & 43.84           \\
\hline
\hline
\end{tabular}
\end{center}
\end{table*}
coincides again with 3C111. 
Note the close position of sources n.4 and n.5, which suggests a likely 
association; in fact source n.5 is identified in NED with a symmetric double 
so that n.4 could be part of this source. Except for 3C111, no X-ray emission is detected at the position of these radio 
emitting sources and again NED reveals that most of them have  steep radio spectra.
The most effective way to look for radio candidates of gamma-ray sources is by searching 
high frequency catalogues which are more likely to isolate flat spectrum sources.
Using the Green Bank Radio Source Catalog at 6 cm (Becker et al. 1991) we found three radio sources inside the 99.9{\%}
EGRET $\gamma$-ray probability contours (see squares in Fig. 2)
having  flux greater than 100 mJy (see Table 3).
By far the brightest one is GB6 J0418+3801 (n.9  in Table 3) which coincides with 3C111. 
The GB radio source n.11 is the only  likely flat 
spectrum object in the error box (apart 3C111): combining the 4.85 GHz measurement with  20 cm data we
obtain a radio spectrum with $\alpha$=-1.09 (S$_{\nu}\propto\nu^{-\alpha}$).
Nevertheless this radio source and the other one (n.10 in Table 3) have no X-ray 
counterpart and are too dim in 
radio to be considered  potentially likely counterparts. \\
To conclude, 3C111 is not only the strongest hard X-ray source in the EGRET error box but 
it is also by far the brightest radio object in the vicinity of 3EG J0416+3650.
\section{BeppoSAX observations of 3C111}
BeppoSAX-NFI pointed at 3C111 in March 1998, for an effective MECS and
PDS exposure time of 69 Ks and 33 Ks respectively.  MECS data have been
downloaded from the BeppoSAX archive and a standard analysis was
performed using SAXDAS software.  The PDS spectra
were instead extracted using the XAS v2.1 package (Chiappetti \& Dal Fiume
1997) which allows a more reliable check of the backgroud fields.  For
the PDS data we selected visibility windows, following the criteria of
No Earth occultation and high voltage stability during the exposure;
in addition observations closest to the South Atlantic Anomaly were
discarded from the analysis.  A careful analysis of the +OFF -OFF
field was also performed in order to check for the presence of
contaminating sources, which if present would provide an uncorrect
background subtraction; no contamination has been found despite the
location of these offset fields in the galactic plane region.
Spectral fits were performed using the XSPEC 11.0.1 software package
and public response matrices as from the 1998 November issue.  MECS PI
(Pulse Invariant) channels were rebinned in order to sample the
instrument resolution with an accuracy proportional to the count rate.
The PDS data were instead rebinned so as to have logarithmically equal
energy intervals.  The data rebinning is also required to have at
least 20 counts in each bin such that the $\Delta\chi^2$ statistics
could be reliably used.  In the following analysis, we use an absorbed
component to take into account the galactic absorption that in the
direction of 3C111 is $\sim$1.2$\times$ 10$^{22}$ cm$^{-2}$, if we
consider both the molecular and neutral hydrogen column density (Bania
et al. 1991). The quoted errors in the following correspond to 90\% confidence level
for one interesting parameter ($\Delta\chi^2$ =2.71). \\ 3C111 is well
detected in the 2-10 keV energy range at 125$\sigma$ (see Fig. 1 where 
the MECS image has been superimposed on the EGRET countours). At higher
energy, in the PDS regime, a 10.5$\sigma$ detection is also
measured.\\ A fit to the combined MECS/PDS data sets provides a value
of 0.81$^{+0.13}_{-0.11}$ for the cross calibration constant between
the two instruments, fully compatible within the errors with the
nominal range expected (Fiore, Guainazzi \& Grandi 1998).
The broad band (2-200 keV) spectrum of the source provides a power law
with a hard photon index of $\Gamma$=1.64$^{+0.056}_{-0.056}$.
Neither a reflection component nor an iron line is required by the
data and  there is no evidence in the data for absorption in
excess to the galactic value.
Furthermore  the BeppoSAX spectrum does not show evidence for a high energy break up to about
200 keV. Overall we conclude that the X-ray  
power law is free from extra features (absorption, reflection, line, cutoff), i.e.  is blazar like.
The fluxes corrected for absorption are respectively 2.7  $\times$ 10$^{-11}$  and 9 $\times$ 10$^{-11}$ erg cm$^{-2}$ s$^{-1}$ for
the 2-10 and 15-200 keV band; the corresponding luminosities 
(at the reported redshift and assuming H$_{\circ}$= 75, $\lambda$$_{\circ}$=0.7) 
are  respectively 1.3 $\times$ 10$^{44}$ and 4.2 $\times$ 10$^{44}$  erg s$^{-1}$.
\section{3C111 as a misaligned blazar}
3C111 is a near (z=0.0485) Fanaroff-Riley (FR) type II  radio source and  displays several small-scale features
 characteristic of a highly active nucleus.\\
The bright and variable central component is coincident with a broad line galaxy classified as 
a Seyfert 1 (NED). Few morphologically details are known of the optical galaxy since it is behind a heavily obscured region identified with the galactic dark cloud complex Taurus B.
(Ungerer et al. 1985).  It is a  very small
and elliptical-like galaxy strongly dominated by a bright point like nucleus 
with a very faint underlying fuzz.\\ 
The radio morphology is typical of FR II objects with a double-lobe/single 
jet structure (Linfield and Perley 1984). On kiloparsec scales, the one sided highly collimated jet is 
seen emerging from the core at a position angle of  $\sim$63$^{\circ}$ and leading 
to the spot at the edge of one of the two lobes; there is probably an active but undetected 
counterjet feeding the other lobe. On parsec scales, the jet is much more prominent and 
is roughly aligned with the kpc jet (Linfield 1981). The central component is strongly variable on a time scale 
of a few months: high resolution observations  
at 6 cm  show dramatic changes in the structure of the compact emission region implying superluminal behaviour
(Preuss et al. 1988, 1990).
The radio spectrum of 3C111 is flat having an index of -0.3 ($\alpha$ $\ge$ -0.5,  S$_{\nu}\propto\nu^{\alpha}$)
from 6 cm to 80 cm 
(Becker, White and Edwards 1991).\\
In the mm band, 3C111 has the strongest compact core of all FR II radio galaxies.
The spectrum between 4 cm and 3 mm is flat or inverted (Bloom et al. 1994).
At times 3C111 displays superluminal structural changes and/or strong mm outbursts. A large outburst,
with flux values $>$10 Jy at 3.3 mm, was discovered with IRAM interferometer in January 1996 (Alef et al. 1998).  
Reich et al. (1993) have shown that hard $\gamma$-ray emission from EGRET blazars is associated with high 
and variable mm flux densities.  During March 1991 and July 1992 (close to the EGRET Viewing Period
respectively VP=0.2+ and VP=31.0)
3C111 has been observed at 3.3 mm wavelenght (Steppe et al. 1993), showing high flux densities (respectively 3.6 Jy and 3.2 Jy).\\
In the far-infrared, 3C111 is a strong emitter detected by IRAS (Golombek et al. 1988). 
2MASS measurements are also available (band J, H, K) from the HEASARC  database.\\ 
In the X-ray band the source has been observed by every major instrument. 
The long term behaviour of the source suggest at least a factor of 5 variation in the 
2-10 keV range (Reynolds et al. 1998), while no significant fluctuations on time scales less than 
an hour have been observed (Eracleous et al. 2000).\\
As for the $\gamma$-ray band, 3C111 has not been detected by COMPTEL (Maisack et al. 1997) placing
2$\sigma$ upper limits on its emission at energies 0.75-1 MeV, 1-3 MeV, 3-10 MeV.
Furthermore, assuming that 3C111 is the counterpart of 3EG J0416+3650, EGRET measurements 
at energies E$>$100 Mev are also available (Hartman et al. 1999).\\ 
To conclude several pieces of evidence point to the idea that 3C111 is a blazar like object:
it is a flat spectrum radio source with  a strong radio core strongly  polarized, it is a strong millimeter emitter,
it is a superluminal source and shows strong variability in every wavebands where it has 
been monitored; furthermore the X-ray spectrum is blazar like lacking all features due 
to reprocessing emission.\\
To confirm this behaviour we have also constructed the source SED
\begin{figure}
\psfig{figure=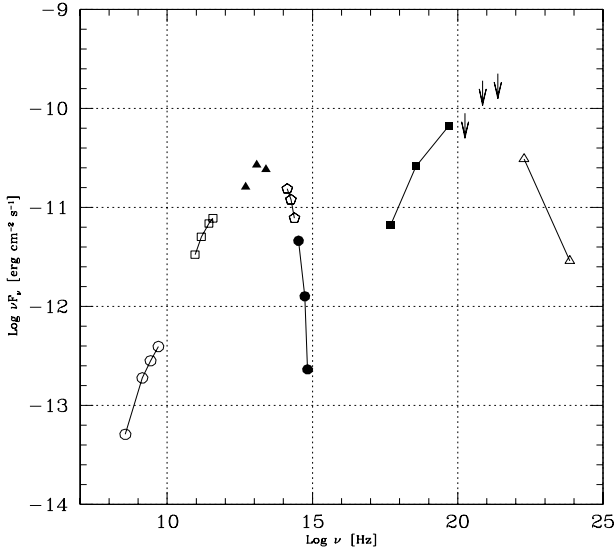,height=9.0cm,width=9.5cm}
\caption{The observed SED of 3C111. We shown here all published data.
Open circles are radio measurements, open squares are millimeter measurements, filled triangles 
are IRAS data,
open pentagons are 2MASS measurements (bands J,H,K),  
 filled circles are optical measurements (bands I,V,B), filled squares are our BeppoSAX measurements,
 open triangles are EGRET measurementsa and arrows are 2$\sigma$ COMPTEL upper limits (0.75-1 MeV, 
1-3MeV, 3-10 MeV).}
\end{figure}
\begin{figure}
\psfig{figure=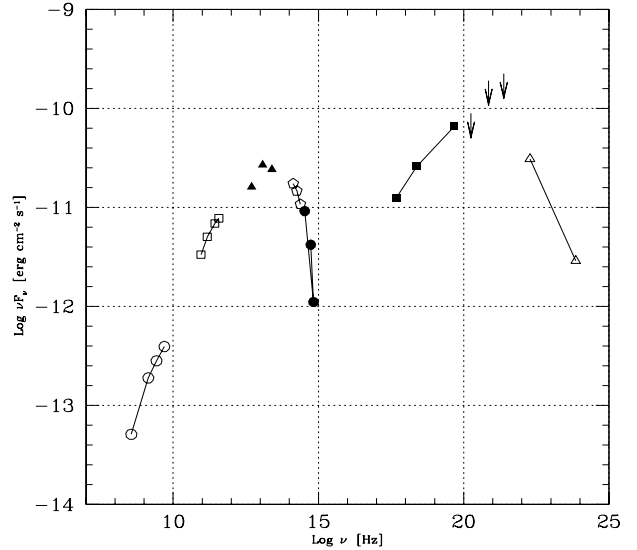,height=9.0cm,width=9.5cm}
\caption{The corrected SED of 3C111. Optical and far-infrared data points(filled circles and open pentagons,
respectively)have been dereddened with A$_V$=1.3, while the X-ray point at 2 keV  has been corrected
for N$_H$=1.2$\times$10$^{22}$ cm$^{-2}$.}
\end{figure}
collecting, from the literature and from the BeppoSAX results presented in this paper, fluxes spanning the widest range of frequencies, 
from the radio to the gamma-ray band.
These data produce the observed SED shown in Fig. 3, which appears
to be composed of two broad peaks, their maxima occur in spectral bands which are essentially
unaffected by dust/gas absorption (respectively far-infrared and soft gamma-ray range).
However absorption must be taken into consideration, expecially in view of the location
of 3C111 behind the Taurus cloud complex: absortpion would indeed naturally steepen the
Infrared to UV and soft to hard X-ray slopes of the continuum emission thus artificially producing a two
peak SED. While the correction in X-ray is easily provided by the estimate of the column density, 
the extinction in optical is more complex to evaluate. The visual extinction map provided by Ungerer et 
co-workers (1985) over an area of 1.4$^{\circ}$x1.2$^{\circ}$ reveals the existence of a large condensation
with small clumps of high visual extinction; however the part of the cloud just in front of 3C111 is not
the densest and therefore in this direction the extinction is moderate, A$_V$ being only 1.3 magnitude.
Correcting both optical/near-infrared  and the  X-ray  
band for the appropriate extimated extinction/absorption we obtain Fig. 4.
The main effect of this procedure is to reduce substantially the depth of the minimum
at $\nu$$\simeq$10$^{15}$ Hz. Even after this correction, the two broad peaks seen in Fig. 4 are 
still present and by analogy with blazars, can be  attribuited to  non thermal
synchroton emission and to inverse Compton scattering of softer photons.
Furthermore the X/gamma radiation 
completely dominates the radiative output: the SED of 3C111 thus intriguingly resembles those
of Compton dominated blazars.
These general features are quite easily explained in the layer-spine scenario of Ghisellini, 
Tavecchio and Chiaberge (2004).
\section{Probability of association}
The error box of 3EG J0416+3650 does not contain any obvious galactic counterparts.
The variability of the EGRET source excludes a pulsar as a potential candidate as well 
as a  supenova remnants (SN). Pulsar Wind Nebulae can be strong and 
variable gamma-ray emitters (Nolan et al 2003), but using the HEASARC database we found none  
inside the EGRET  error box; furthermore PWN are expected to be both radio and X-ray 
emitting objects (Roberts, Gaensler and Romani 2002) while none of the objects we found within the 99.9$\%$ 
contours is both a radio and an X-ray source except for 3C111. 
We are therefore left with the possibility of an association with 
a galactic object such as a microquasars or a Wolf Rayet star, but also in this case no object of this type has been found
inside the EGRET error box (HEASARC). \\
Bearing this in mind, the only other possible counterpart is an extragalactic object, with characteristics similar to
those observed in blazars. 3C111 is fully compatible with this scenario.
In the light of our findings we can revaluate the a posteriori probability of association following
Mattox et al. (1997, 2001). The factors that we used for this calculation are the radio 
flux (6.637 Jy) and the spectral index
(-0.3) of 3C111, the 95\% error radius of the EGRET source (0$^{\circ}$.66), the distance of 3C111 
from the center of the EGRET error box (1$^{\circ}$.27) and the mean distance between radio sources 
that are at least as strong and at least as flat as 3C111 ($\sim$43$^{\circ}$).
The value obtained ($\sim$0.05 or $\sim$5\%) is not high enough to consider the association highly probable
because of the location of 3C111 at the border of the EGRET 
error box. However, the location outside  the 99$\%$  
$\gamma$-ray probability countour (but inside the  99.9$\%$ one) 
should  not be taken as a strong element to exclude 3C111
as a possible counterpart of the EGRET source.
As pointed out by Thompson et al. (1995) and Hartman et al. (1999), the position uncertainty region provided in the EGRET catalogues are statistical only: 
a 'generous' allowance  for an additional systematic position error due to  
inaccuracies in the Galactic diffuse radiation model  and potential source
confusion is recommended to the user, particularly for those EGRET sources located along the galactic plane.
Therefore it is clearly
worth  searching for counterparts even outside the 99\% error box.
\section{Conclusions}
From analysis of archived radio, millimeter, infrared, optical and X/gamma-ray data,  
we conclude that  3EG J0416+3650 is likely to be associated with the radio galaxy 3C111; furthermore
we show that this source can be interpreted as a misoriented blazar, mainly suggested 
on the basis of its SED and multiwavelenght characteristics. Bearing this in mind, 3C111 could be the third radio 
galaxy detected by EGRET, after Centaurus A and possibly NGC6251.
This association strongly supports the scenario that radio galaxies, not only blazars,
can be high energy $\gamma$-ray emitters, as suggested by Ghisellini et al. (2004). This could have 
important consequences in opening the search for counterparts to a more populated class of objects
and in providing a new perspective in the relation between blazars and radio galaxies.

\begin{acknowledgements}
This research has made use of SAXDAS linearized and cleaned MECS event files produced at
the BeppoSAX Science Data Center and cleaned events PDS files produced by using XAS package. 
It halso also made use of data obtained from the HEASARC, SIMBAD and NED database.
This research has been supported by Southampton University School of Astronomy and Astrophysics.
\end{acknowledgements}

\end{document}